\documentclass[prb,showpacs]{revtex4}% Physical Review B
\usepackage{amsfonts}
\usepackage{amsmath}
\usepackage{amssymb}
\usepackage{graphicx}%
\begin{document}

\title{Optical properties of polaronic excitons in stacked
quantum dots}
\author{V. N. Gladilin, S. N. Klimin, V. M. Fomin, and J. T. Devreese}
\affiliation{Theoretische Fysica van de Vaste Stoffen, Departement Natuurkunde, Universiteit Antwerpen,
Universiteitsplein 1, B-2610 Antwerpen, Belgium}

\begin{abstract}
We present a theoretical investigation of the optical properties of
polaronic excitons in stacked self-assembled quantum dots, which is
based on the non-adiabatic approach.  A parallelepiped-shaped quantum dot is
considered as a model for a self-assembled quantum dot in a stack. The
exciton-phonon interaction is taken into account for all phonon modes
specific for these quantum dots (bulk-like, half-space and interface
phonons). We show that the coupling between stacked quantum dots can lead to a strong enhancement of the optical absorption in the spectral ranges characteristic for phonon satellites.

\end{abstract}
\date{\today}

\maketitle

Non-adiabaticity is an inherent property of exciton-phonon systems in
various quantum-dot structures.  Non-adiabaticity \emph{drastically enhances the
efficiency of the exciton-phonon interaction}. The effects of
non-adiabaticity are important to interpret the surprisingly high
intensities of the phonon `sidebands' observed in the optical absorption,
the photoluminescence and the Raman spectra of quantum dots, in particular, an enhancement of these intensities with decreasing the quantum-dot size (see, e.~g., Refs.~\onlinecite{JH96,cap2000}).  
Deviations of intensities of the phonon-peak sidebands, observed in some experimental optical spectra, from the Franck-Condon progression, which is prescribed by the commonly
used adiabatic approximation, find a natural explanation within our
non-adiabatic approach \cite{12,nanot,PSS03,raman2002}. Recently, stacked
quantum dots have received increasing attention (see, e.~g.,
Refs.~\onlinecite{Schmidt97,Luyken99,NN1,pi01,Bayer01,NN4,NN5}) due to the possibility to finely control their energy spectra. This makes stacked quantum dots very promising for future nanodevices \cite{NN4,NN5}. In the present work,
the non-adiabatic approach is applied to stacked InAs/GaAs quantum dots,
which reveal a richer structure of phonon and exciton
spectra in comparison with those for a single quantum dot.

In order to model coupled self-assembled InAs/GaAs quantum dots, we
consider a stack of $N$ parallelepiped-shaped quantum dots with sizes $l_{x},$
$l_{y},$ $l_{z}$ and interdot distance $d$ along the $z$-axis.
Within the present approach, the lateral sizes ($l_{x},$ $l_{y}$) of each
quantum dot in a stack are supposed to be much larger than its size
$l_{z}$ along the growth axis. The stack is a
system of $\left(  2N+1\right)  $ layers ($n=1,\ldots2N+1$) with parameters%
\[
\left\{
\begin{array}
[c]{ll}%
l_{n}=l_{z},\;\varepsilon_{n}=\varepsilon_{\mathrm{InAs}} & \mathrm{for}%
\;n=2,4,\ldots,2N;\\
l_{n}=d,\;\varepsilon_{n}=\varepsilon_{\mathrm{GaAs}} & \mathrm{for}%
\;n=3,5,\ldots,2N-1;\\
l_{n}\rightarrow\infty,\;\varepsilon_{n}=\varepsilon_{\mathrm{GaAs}} &
\mathrm{for}\;n=1,2N+1.
\end{array}
\right.
\]

The bulk-like optical-phonon frequencies in InAs and GaAs layers of the
stacked InAs/GaAs quantum dots coincide with the LO-phonon frequencies in InAs
and GaAs, respectively. The interface frequencies belong to the stacked
quantum
dots as a whole and satisfy the dispersion equation%
\begin{equation}
\det\left\|  a_{kn}\left(  \omega\right)  \right\|  =0\quad\left(
k,n=1,\ldots2N\right)  ,\label{DispEq}%
\end{equation}
where $a_{kn}\left(  \omega\right)  $ is the dynamic matrix of the
interface
vibrations with the matrix elements%
\begin{equation}
\left\{
\begin{array}
[c]{l}%
a_{nn}\left(  \omega\right)  =\varepsilon_{n}\left(  \omega\right)  \coth
q_{\parallel}l_{n}+\varepsilon_{n+1}\left(  \omega\right)  \coth
q_{\parallel
}l_{n+1},\\
a_{n,n-1}\left(  \omega\right)  =a_{n-1,n}\left(  \omega\right)
=-\varepsilon_{n}\left(  \omega\right)  /\sinh q_{\parallel}l_{n};
\end{array}
\right.  \label{MatrixR}%
\end{equation}
all other matrix elements are equal to zero.

\begin{figure}
\centering \includegraphics[scale=0.7]{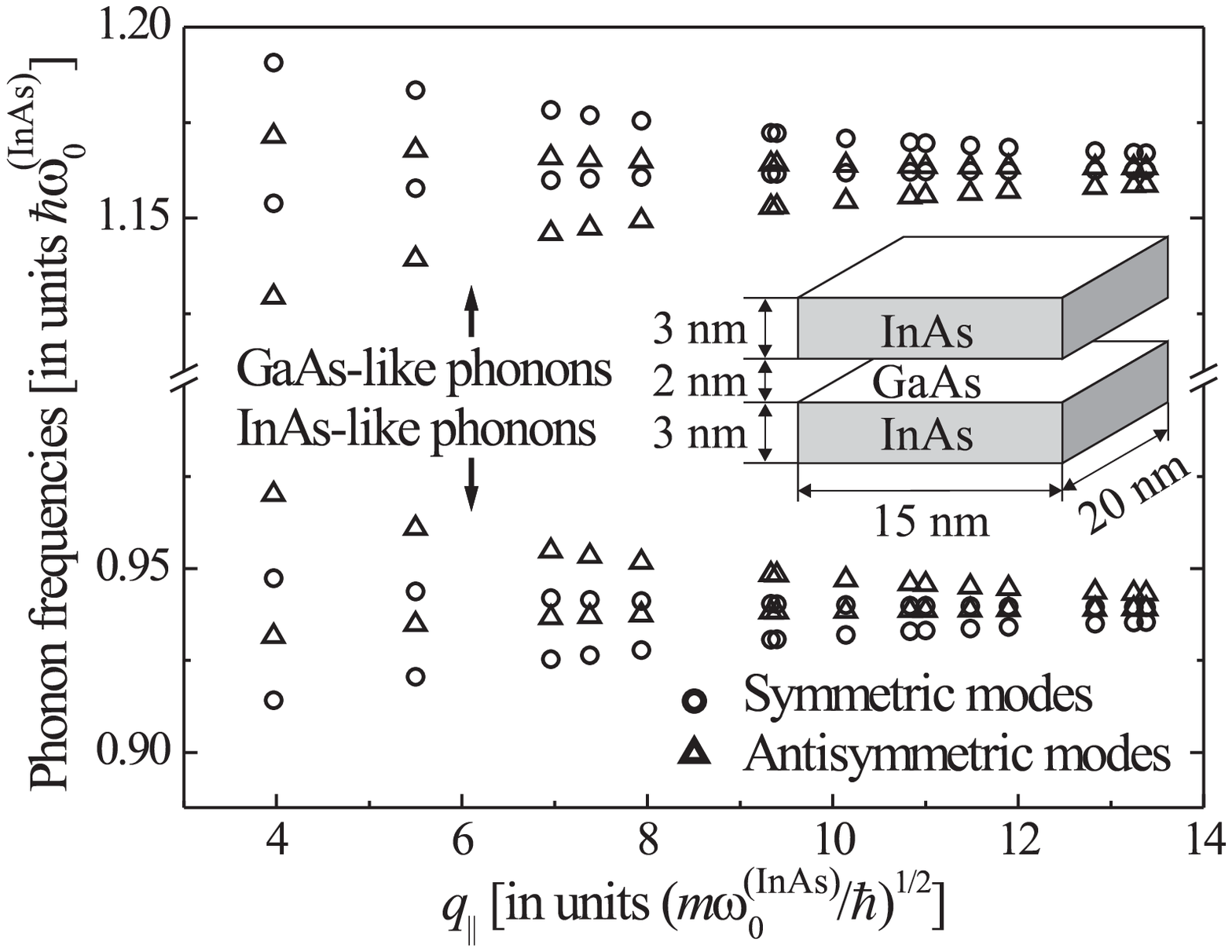}%
\caption{Interface-phonon frequencies for two stacked parallelepiped-shaped InAs/GaAs quantum dots. $\omega_0^{\rm (InAs)}$ is the frequency of LO phonons in InAs at the center of the Brillouin zone.
}
\label{fig-stack1}
\end{figure}

In Fig. {\ref{fig-stack1}, typical interface-phonon spectra are
represented for stacked InAs/GaAs quantum dots formed by two InAs
parallelepipeds. The frequencies are plotted as a function of the in-plane
wave number $q_{\parallel}$, which takes discrete values due to the
quantization of the phonons in the $xy$-plane. In a stack of $N$ quantum dots,
each interface-phonon frequency of a single quantum dot splits into $N$
branches. The splitting of the interface-phonon frequencies is due to the
electrostatic interaction between the optical polar vibrations of the different quantum dots. }

These features of the optical-phonon spectrum of stacked quantum
dots are manifested in their optical properties. We calculate the optical absorption spectrum of 
polaronic excitons in stacked quantum dots starting from the Kubo formula. Within the non-adiabatic approach~\cite{12} the following
expression results for the linear coefficient of the optical absorption by
the exciton-phonon system in a quantum-dot structure:
\begin{equation}
 \alpha(\Omega)\propto\mbox{Re}  \sum\limits_{\beta,\beta^{\prime}}d_{\beta
}^{*} d_{\beta^{\prime}}^{} \int\limits_{0}^{\infty}dt\, e^{i(\Omega-
\Omega_{\beta}+i0^+)t} \left\langle \beta\left|  \bar U(t) \right|
\beta^{\prime}\right\rangle , \label{A1}
\end{equation}
where $\Omega$ is the frequency of the incident light, $d_{\beta} $ and $\Omega_{\beta}$
are, respectively, the electric dipole matrix element and the Franck-Condon frequency
of a transition between the exciton vacuum state and the one-exciton state
$|\beta\rangle$. Exciton states in stacked InAs/GaAs quantum dots are determined using an exact diagonalization of the exciton Hamiltonian with simple parabolic valence and conduction bands within a finite-dimensional basis of the
electron-hole states. The evolution operator averaged over the phonon ensemble,
$\bar U(t)$, is 
\begin{align}
  \bar U(t) &=\mathrm{T} \exp\Biggl \{  -\frac{1}{\hbar^{2}} \sum
\limits_{\lambda}\int\limits_{0}^{t} dt_{1} \int\limits_{0}^{t_{1}} dt_{2}
\left[ \frac{\mbox{e}
^{-i\omega_{\lambda} \left(  t_{1}-t_{2}\right)}}{1-y_\lambda}\right.\Biggr.
\nonumber\\
&  \Biggl.\left. \times  \gamma_{\lambda}^{}
(t_{1})\gamma^{\dag}_{\lambda}(t_{2}) + \frac{y_\lambda \mbox{e}
^{i\omega_{\lambda} \left(  t_{1}-t_{2}\right)}}{1-y_\lambda} \gamma^{\dag}_{\lambda
}(t_{1})\gamma_{\lambda}^{}(t_{2})\right]  \Biggr \}  . \label{A3}%
\end{align}
In Eq.~(\ref{A3}), $\mathrm{T}$ is the time ordering operator, the index $\lambda$ labels the phonon modes specific for the
quantum-dot structure under consideration, $\omega_{\lambda}$ are phonon
frequencies, $\gamma_{\lambda}(t)$ are the exciton-phonon interaction
amplitudes in the interaction representation, and \linebreak $y_{\lambda} =
\mbox{exp}  \left(-  \hbar\omega_{\lambda}/ k_{B} T\right)  $.

Within the adiabatic approximation, which has been widely used to calculate the
optical spectra of quantum dots, non-diagonal matrix elements of the
exciton-phonon interaction are neglected when calculating $\alpha(\Omega)$ as given by Eq.~(\ref{A1}) with Eq.~(\ref{A3}). In the adiabatic
approach~\cite{8,9} one supposes that (i) both the initial and the final states of
a quantum transition are non-degenerate, (ii) the energy differences between
the exciton states are much larger than the phonon energies. It has been
shown in Refs.~\onlinecite{12,nanot,PSS03,raman2002} that these conditions are often violated for optical transitions in small
quantum dots, which have sizes less than the bulk exciton radius. In other words, the exciton-phonon system in a quantum dot can
be essentially \emph{non-adiabatic}. The polaron interaction for an exciton in
a degenerate state results in \emph{internal non-adiabaticity} (``the proper
Jahn--Teller effect''), while the existence of exciton levels separated by an
energy comparable with the LO-phonon energy leads to \emph{external
non-adiabaticity} (``the pseudo Jahn--Teller effect'').

In Ref. \onlinecite{12}, a method was proposed to calculate the absorption
spectrum given by Eqs. (\ref{A1}) and (\ref{A3}) taking into account the
effect of non-adiabaticity on the probabilities of phonon-assisted optical
transitions. The key step is the calculation of the matrix elements of the
evolution operator $ \langle \beta\left| \bar U(t) \right|\beta^\prime
\rangle $. In order to describe the effect of non-adiabaticity both on
the intensities and on the positions of the absorption peaks, a diagrammatic approach
can be used. When calculating these matrix elements we take into account that in
a quantum dot, due to the absence of momentum conservation, the product
$\langle \beta_1 | \gamma_{\lambda}^{} |\beta_2 \rangle \langle \beta_2 |
\gamma_{\lambda}^{*} |\beta_3\rangle$ can be non-zero for
$\beta_1\neq\beta_3$, as distinct from the bulk case. Consequently,
the evolution operator is in
general non-diagonal in the basis of one-exciton wavefunctions $ |\beta
\rangle$. For the absorption coefficient we obtain
\begin{align}
\alpha (\Omega )&\propto -{\rm Im} \sum\limits_\beta |d_{\beta}|^2
G_{\beta }\left(\Omega +i0^+\right)-{\rm Im} \sum\limits_{\beta,
\beta^{\prime} } d_{\beta} d^*_{\beta^{\prime}}
\nonumber \\
&\times \left[Q^{(1)}_{\beta\beta^{\prime}}(\Omega+i0^+)+
Q^{(2)}_{\beta\beta^{\prime}} (\Omega+i0^+)\right], \label{n1}
\end{align}
where
\begin{eqnarray}
G_{\beta }\left(\Omega \right)&&=
\sum\limits_{\{j_\lambda =-\infty\} }^{\{\infty\}}
\frac{
C_{\{j_\lambda\} \beta }^{(+)} }
{ A_{ \beta } \left(\Omega-\sum\limits_{\lambda} j_\lambda\omega_\lambda \right)},
\label{n2}\end{eqnarray}
\begin{eqnarray}
A_{ \beta } \left(\Omega \right)=
\Omega-\Omega_\beta+\sum\limits_{\lambda} S_{\lambda,\beta}\omega_\lambda -
\Sigma_\beta^{(1)} \left(\Omega \right)- \Sigma_\beta^{(2)}\left(\Omega \right),
\label{n2aa}\end{eqnarray}
\begin{align}
C_{\{j_\lambda\} \beta }^{(\pm)} &= \prod\limits_{\lambda} (\pm 1)^{
j_\lambda } {\rm exp}\left[ \mp (2 \bar n_\lambda +1)S_{\lambda\beta}
+\frac{ j_\lambda \hbar\omega_\lambda }{2 k_B T} \right] \nonumber\\
&\times I_{|j_\lambda|} \left( {S_{\lambda\beta}} \left[2 \sinh
\left(\frac{\hbar\omega_\lambda }{2 k_B T} \right)\right]^{-1} \right).
\label{n3}\end{align} 
$I_n(x)$ is a modified Bessel function of the first kind and
$S_{\lambda\beta}$ is the Huang-Rhys parameter, which is related to the
interaction of the exciton in the state $\beta$ with phonons of the
$\lambda$-th mode:
\begin{eqnarray}
S_{\lambda\beta}=
\left|\frac{\langle \beta | \gamma_\lambda |\beta \rangle}{\hbar\omega_\lambda}\right|^2.
\label{n2a}\end{eqnarray}

The self-energy terms $\Sigma_\beta^{(1)} \left(\Omega\right)$ and
$\Sigma_\beta^{(2)} \left(\Omega\right)$ in Eq.~(\ref{n2aa}) are obtained
by  summing diagrams, which describe one- and two-phonon non-adiabatic
contributions:
\begin{eqnarray}
\Sigma_\beta^{(1)} \left(\Omega\right)=
\sum\limits_{j=\pm 1}
\sum\limits_{\lambda, \beta_1}
F_{\beta \beta_1}\left(\Omega-j\omega_\lambda\right)
M_{\lambda \beta \beta_1 \beta_1 \beta}^{(j)}
\label{n5}\end{eqnarray}
and
\begin{align}
\Sigma_\beta^{(2)} \left(\Omega\right)& = \sum\limits_{j_1,j_2=\pm 1} \sum\limits_{\lambda_1,
\lambda_2}
\sum\limits_{\beta_1, \beta_2, \beta_3}
F_{\beta \beta_1}\left(\Omega-j_1\omega_{\lambda_1}\right)
\nonumber\\
&\times
F_{\beta \beta_3}\left(\Omega-j_2\omega_{\lambda_2}\right)
\left[
F_{\beta \beta_2}\left(\Omega-j_1\omega_{\lambda_1}-j_2 \omega_{\lambda_2}\right)
\phantom{M_{\lambda_2 \beta_1 \beta_2 \beta_3\beta }^{(j_2)}}
\right.
\nonumber\\
&\times M_{\lambda_1 \beta \beta_1 \beta_2 \beta_3}^{(j_1)} M_{\lambda_2
\beta_1 \beta_2 \beta_3\beta }^{(j_2)}+F_{\beta
\beta_2}\left(\Omega\right)
\nonumber\\
&\times\left.
M_{\lambda_1 \beta \beta_1 \beta_1 \beta_2}^{(j_1)}
M_{\lambda_2 \beta_2 \beta_3 \beta_3 \beta}^{(j_2)}
\left(1-\delta_{\beta_2\beta}\right)
\right]
\label{n7}
\end{align}
where
\begin{equation}
F_{\beta \beta_1}\left(\Omega \right)=
\sum\limits_{\{j_\lambda =-\infty\} }^{\{\infty\}}
C_{\{j_\lambda\} \beta }^{(-)}
 G_{ \beta_1 } \left(\Omega-\sum\limits_{\lambda}j_\lambda\omega_\lambda \right),
\label{n2bb}
\end{equation}
\begin{align}
M_{\lambda \beta_1 \beta_2 \beta_3 \beta_4}^{(j)}=
m_{\lambda \beta_1 \beta_2 \beta_3 \beta_4}^{(j)}-
m_{\lambda \beta_1 \beta_1\beta_1\beta_1}^{(j)}
\delta_{\beta_1\beta_2}\delta_{\beta_3\beta_4},
\label{nmmm}
\end{align}
\begin{eqnarray}
m_{\lambda \beta_1 \beta_2 \beta_3 \beta_4}^{(j)}= \frac{j\langle
\beta_{2-j} | \gamma_\lambda^{} |\beta_{3-j} \rangle \langle \beta_{2+j} |
\gamma_\lambda^{\dag} |\beta_{3+j}
\rangle}{\hbar^2\left(1-y_\lambda^j\right)}. \label{mplus}
\end{eqnarray}
The functions $Q^{(1)}_{\beta\beta^{\prime}}(\Omega)$ and
$Q^{(2)}_{\beta\beta^{\prime}}(\Omega)$ in Eq.~(\ref{n1}), which describe
contributions of one- and two-phonon processes to non-diagonal matrix
elements of the evolution operator, take the form
\begin{align}
Q^{(1)}_{\beta\beta^{\prime}}(\Omega)&=G_{\beta}\left(\Omega \right)G_{\beta^\prime}\left(\Omega \right)
\left(1-\delta_{\beta\beta^\prime}\right)
\nonumber\\
&\times
\sum\limits_{j=\pm 1}
\sum\limits_{\lambda,\beta_1}
G_{\beta_1}\left(\Omega-j\omega_\lambda\right)
m_{\lambda \beta \beta_1 \beta_1 \beta^\prime}^{(j)},
\label{n8}
\end{align}
\begin{align}
Q^{(2)}_{\beta\beta^{\prime}} (\Omega) &= G_{\beta}\left(\Omega\right)
G_{\beta^\prime}\left(\Omega\right)
\left(1-\delta_{\beta\beta^\prime}\right)
\nonumber \\
&\times
\sum\limits_{j_1,j_2=\pm 1}
\sum\limits_{\lambda_1,
\lambda_2}
\sum\limits_{\beta_1, \beta_2, \beta_3}
G_{\beta_1}\left(\Omega-j_1\omega_{\lambda_1}\right)
\nonumber\\
&\times
G_{\beta_3}\left(\Omega-j_2\omega_{\lambda_2}\right)
\left\{
G_{\beta_2}\left(\Omega-j_1\omega_{\lambda_1}- j_2\omega_{\lambda_2}\right)
\phantom{M_{\lambda_2 \beta_1 \beta_2 \beta_3\beta }^{(j_2)}}
\right.
\nonumber\\
&\times \left[ \left(1-\delta_{\beta_3\beta_1}\right) m_{\lambda_1 \beta
\beta_1 \beta_3 \beta^\prime}^{(j_1)} m_{\lambda_2 \beta_1 \beta_2 \beta_2
\beta_3}^{(j_2)} \right. \nonumber\\ &+ \left. m_{\lambda_1 \beta \beta_1
\beta_2 \beta_3}^{(j_1)} m_{\lambda_2 \beta_1 \beta_2 \beta_3
\beta^\prime}^{(j_2)} \right] +G_{\beta_2}\left(\Omega\right)
\left(1-\delta_{\beta_2\beta}\right) \nonumber\\ &\times\left.
\left(1-\delta_{\beta_2\beta^\prime}\right) m_{\lambda_1 \beta \beta_1
\beta_1 \beta_2}^{(j_1)}m_{\lambda_2 \beta_2 \beta_3 \beta_3
\beta^\prime}^{(j_2)}. \right\} \label{n9}\end{align}%
The absorption
spectrum is thus expressed through the functions $
G_{\beta}\left(\Omega\right)$, which in turn are determined by a
closed set of equations (\ref{n2}), (\ref{n2aa}), and
(\ref{n5}) to (\ref{n2bb}).

\begin{figure}
\centering \includegraphics[scale=0.8]
{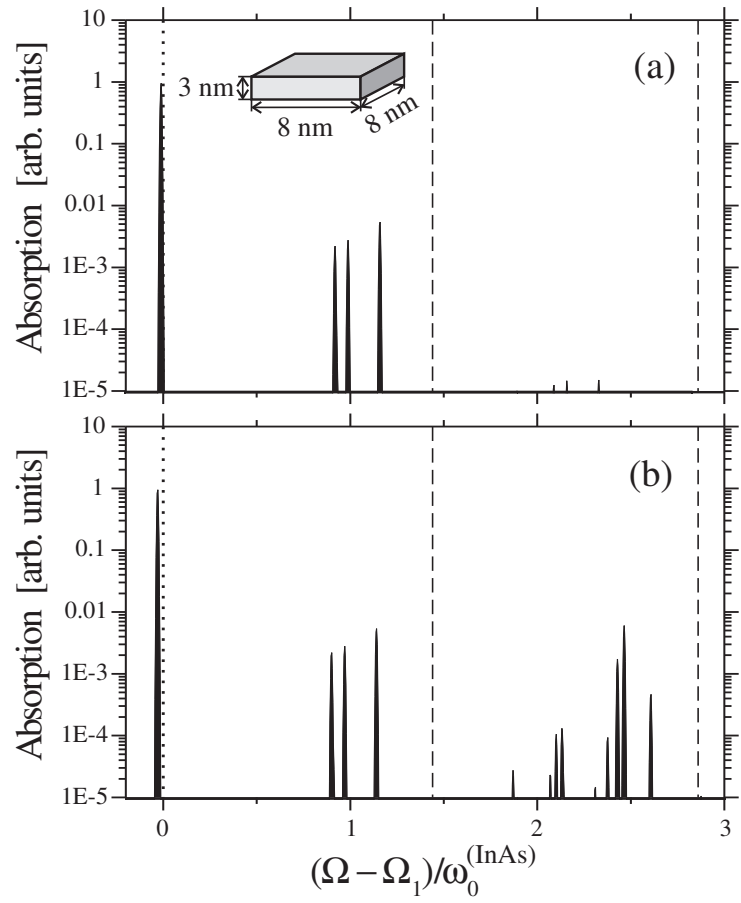}%
\caption{Absorption spectra, calculated for a single quantum dot with the
adiabatic approximation [panel (a)] and with the non-adiabatic approach
[panel (b)]. Optically active and non-active energy levels of a bare exciton
are shown as dotted and dashed lines, respectively. $\Omega_{1}$ is the
transition frequency for the lowest state of a bare exciton.}
\label{fig-stack3}
\end{figure}

\begin{figure}
\centering \includegraphics[scale=0.8]{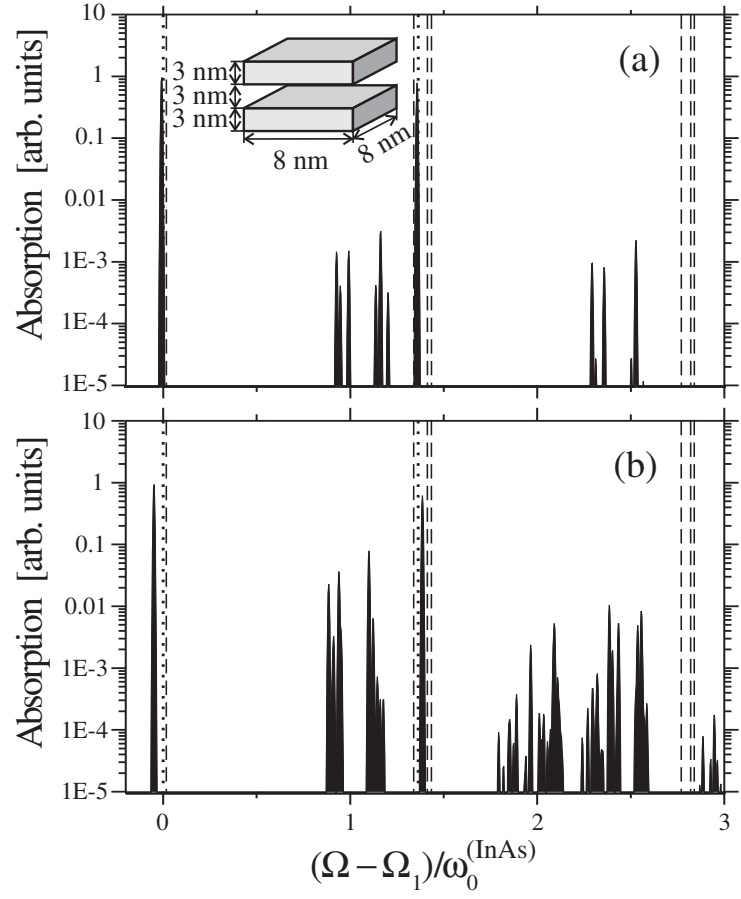}%
\caption{Absorption spectra, calculated for a system of two stacked quantum dots with the adiabatic approximation [panel (a)] and with the non-adiabatic
approach [panel (b)]. Notations are the same as in Fig.~2.}
\label{fig-stack4}
\end{figure}

In Figs.~\ref{fig-stack3} and \ref{fig-stack4} the calculated optical
absorption spectra are shown for a single quantum dot and for a system of two
stacked quantum dots, respectively. The calculations where performed for low temperatures ($\left\{y_\lambda \ll 1\right\}$) when the absorption-line broadening due to the exciton-LO-phonon interaction is negligible. The broadening shown in Figs.~\ref{fig-stack3} and \ref{fig-stack4} is introduced only to enhance visualization. From the comparison of the spectra obtained in
the adiabatic approximation with those resulting from the non-adiabatic approach, the
following effects of non-adiabaticity are revealed.
First, the \textit{polaron shift} of the zero-phonon lines with respect to the
bare-exciton levels is larger in the non-adiabatic approach than in the
adiabatic approximation. Second, there is a strong \textit{increase of
the intensities of the phonon satellites} compared to those given by the adiabatic
approximation. This increase can be by more than two orders of
magnitude. Third, in the optical absorption spectra found within the
non-adiabatic approach, there appear phonon satellites related to
\textit{non-active bare exciton states}.

Fourth, the optical-absorption spectra demonstrate the crucial role of
\textit{non-adiabatic mixing} of different exciton and phonon states in
quantum dots. This results in a rich structure of the absorption
spectrum of the exciton-phonon system \cite{GBFD2001,nanot,PSS03}. For the
stacked quantum dots, this effect is enhanced in the (quasi-)
resonant case, when the exciton-level splitting, caused by the coupling
between quantum dots, is close to a LO phonon energy [see panel (b) in
Fig.~\ref{fig-stack4}]. Similar conclusions about the influence of the
exciton-phonon interaction on the optical spectra of quantum dots have been
recently formulated in Ref.~\onlinecite{VFB2002} for the ``strong coupling
regime'' for excitons and LO phonons. Such a ``strong coupling regime'' is a
particular case of the non-adiabatic mixing related to a (quasi-) resonance, which arises when the spacing between exciton levels is close to the LO phonon energy.

In some cases (see, e.~g., Ref.~\onlinecite{12}) the luminescence spectrum of a quantum dot can be easily derived from its absorption spectrum. E.~g., under the assumption that the distribution function of the states of an exciton coupled to the phonon field, $f(\Omega)$,  depends only on the energy of a state, the luminescence intensity at low temperatures ($\left\{y_\lambda \ll 1\right\}$) can be represented as
\begin{align}
I(\Omega)&\propto
\sum\limits_{K=0}^\infty
\frac{1}{K!}\sum\limits_{\lambda_1,\ldots,\lambda_K}
f\left(\Omega+\sum\limits_{k=0}^K \omega_{\lambda_k}\right)
\nonumber \\
&\times\left.\left(\prod\limits_{k=1}^K \frac{\partial}{
\partial y_{\lambda_k}}\right)
\alpha(\Omega)\right|_{\{y_\lambda \rightarrow 0\}}.
\label{Ial}
\end{align}
Equation~(\ref{Ial}) is applicable, for instance, for thermodynamic equilibrium photoluminescence. In this case the radiative lifetime of an exciton is much larger than the time characteristic of radiationless relaxation between one-exciton states. 

We have calculated the spectra of thermodynamic equilibrium luminescence (not shown here) for quantum dots with parameters indicated above. 
Both for single and for coupled quantum dots, the intensities of phonon satellites in these spectra are significantly smaller than in the absorption spectrum. This is because in quantum dots under consideration the lowest one-exciton energy level (with $\beta=1$) is less affected by the phonon-induced non-adiabatic mixing of states than higher levels. The luminescence spectrum at low temperatures is dominated by transitions from the state with $\beta=1$, while the absorption spectrum contains appreciable contributions due to transitions to higher exciton-phonon states.

Due to non-adiabaticity, multiple absorption peaks appear in the spectral
ranges characteristic for phonon satellites. From the states, which
correspond to these peaks, the system can rapidly relax to the lowest
emitting state. Therefore, in the photoluminescence excitation (PLE)
spectra of quantum dots, pronounced peaks can be expected in spectral
ranges characteristic for phonon satellites. Experimental evidence of
the enhanced phonon-assisted absorption due to non-adiabaticity
has been recently provided by PLE measurements on single self-assembled
InAs/GaAs \cite{lem01} and InGaAs/GaAs \cite{zre01} quantum dots. Our
results, which imply significantly more pronounced phonon satellites in
PLE spectra compared to the luminescence spectra from the lowest
one-exciton state, are in line with the experimental
observations~\cite{lem01,zre01}.

This work has been supported by the GOA BOF UA 2000, I.U.A.P., F.W.O.-V.
projects G.0274.01N, G.0435.03, the W.O.G. WO.025.99N (Belgium) and the
European Commission GROWTH Programme, NANOMAT project, contract No.
G5RD-CT-2001-00545.

\newpage

~

\newpage

~

\newpage

~

\end{document}